\documentclass{iaus}
\usepackage{graphicx}

\newcommand{\Ni}{$^{56}$Ni}
\newcommand{\Fe}{$^{56}$Fe}
\newcommand{\Co}{$^{56}$Co}
\newcommand{\hei}{He$^+$}
\newcommand{\hhh}{H$_3^+$}
\newcommand{\Ms}{$M_{\odot}$}
\newcommand{\gam}{$\gamma$-rays}

\title[JD 11.~~Molecules in supernova ejecta] 
{Molecules in supernova ejecta}

\author[Isabelle Cherchneff \& Arkaprabha Sarangi]   
{Isabelle Cherchneff \& Arkaprabha Sarangi}

\affiliation{Departement Physik, Universit{\"a}t Basel\\ Klingelbergstrasse 82,
CH-4056, Basel, Switzerland\\ email: {\tt isabelle.cherchneff@unibas.ch} \\[\affilskip]}

\pubyear{2011}
\volume{280}  
\pagerange{119--126}
\jname{The Molecular Universe}
\editors{J. Cernicharo \& R. Bachiller, eds.}
\begin{document}

\maketitle

\begin{abstract}
The first molecules detected at infrared wavelengths in the ejecta of a Type II supernova, namely SN1987A, consisted of CO and SiO. Since then, confirmation of the formation of these two species in several other supernovae a few hundred days after explosion has been obtained. However, supernova environments appear to hamper the synthesis of large, complex species due to the lack of microscopically-mixed hydrogen deep in supernova cores. Because these environments also form carbon and silicate dust, it is of importance to understand the role played by molecules in the depletion of elements and how chemical species get incorporated into dust grains. In the present paper, we review our current knowledge of the molecular component of supernova ejecta, and present new trends and results on the synthesis of molecules in these harsh, explosive events.

\keywords{supernovae: general, molecular processes, astrochemistry}
\end{abstract}

\firstsection 
\section{Introduction}

Type II supernovae (SNe) are the end stage of the evolution of massive stars with initial masses on the Zero Age Main Sequence greater than 8 \Ms. These explosive events deposit a tremendous energy to the surroundings with typical explosion energy of 10$^{51}$ ergs per event. The explosion releases a burst of neutrons and synthesised elements heavier than iron through rapid neutron capture (i.e, r-process). The explosion shock wave crosses the helium core, induces a reverse shock at the interface between the heavy elements core and the hydrogen envelope of the stellar progenitor, and reprocesses the ejecta a few hours after explosion. This reverse shock triggers the onset of Raleigh-Taylor instabilities and results in macroscopic mixing between the various core layers. The nebular phase of the SN is reached after $\sim$ 100 days when the luminosity is sustained essentially by the radioactive decay of \Co. Assuming \gam~full trapping, the bolometric light curve declines according to the mean lifetime of \Co~decay into \Fe~($t_{56} = 111.3$ days). In view of this very harsh environment, the detection of molecules in the nebular phase a few months following the explosion of a SN came as a surprise. The locus of the first detection of molecules and dust in a SN event shortly after outburst was the Type II supernova SN1987A in the Large Magellanic Cloud. Indeed, the fundamental band $\Delta v = 1$ of carbon monoxide, CO, at 4.65 $\mu$m was observed between day 135 and day 260 (\cite[Catchpole et al. 1988)]{cat88}, while the CO first overtone transition $\Delta v = 2$ in the region of 2.3 $\mu$m was detected at day 100 after explosion \cite[(Danziger et al. 1988)]{danz88}. Excess emission around $\sim$ 9 $\mu$m was attributed to the fundamental band $\Delta v = 1$ of silicon monoxide, SiO, and was observed as early as day 160 (\cite[Aitken et al. 1988]{ait88}). Other molecular candidates were proposed as carriers of some unidentified bands in the IR spectrum and included CS (3.88 $\mu$m band), H3$^+$ (3.4 $\mu$m and 3.53 $\mu$m bands) and CO$^+$ (2.26 $\mu$m band). However, claims for their presence were not confirmed by theoretical models. 

Since then, carbon and silicon monoxide ro-vibrational emission bands have been observed in other Type II SNe as well and appear to be the only molecules detected in SN ejecta to date. This restricted number of observed species may point to the specific environment of SN ejecta characterised by harsh conditions, the presence of helium ions, and the lack of microscopically-mixed hydrogen. This hydrogen-deprived medium fosters a 'poor' chemistry. Indeed, because of the absence of radicals like hydrides (OH, CH etc.), a fast, hot chemistry that builds up larger, more complex species (e.g., hydrocarbons, water or carbon dioxide) cannot proceed, as it is the case in other circumstellar environments [e.g., AGB stellar winds \cite[(Cherchneff 2006)]{cher06}]. The chemical species thus formed are essentially simple diatomic molecules like CO and SiO.

\section{Observations of molecules and existing models}
The detection in the near IR of the first overtone band of CO has been positive in several Type II SNe at epochs ranging from 90 to 130 days after explosion. Along with the first overtone, the red wing of the fundamental band $\Delta v = 1$ at 4.65 $\mu$m is also detected in the mid-IR (e.g., Kotak et al. 2005). Models of the CO bands indicate conditions in the ejecta close to those provided by explosion models $\sim$ 130 days post-outburst. i.e., gas temperature close to 5000 K. The CO molecule is always present in SN ejecta some 100 days after outburst, pointing to an efficient chemical pathway to its synthesis at high temperatures. In SN1987A, the CO first overtone band flux peaks around 180 days before declining by several orders of magnitude up to day 600 where it became unobservable \cite[(Danziger et al. 1989)]{danz89}. A similar behaviour is observed in other SNe, for example, SN2005af (Kotak et al. 2005). This decrease can be ascribed either to a change in CO mass with time or to different excitation conditions owing to the change in gas temperature, or both. In SN1987A, the fundamental band of CO is also observed to decrease in absolute strength with time but at a much smaller rate than the first overtone band \cite[(Danziger et al. 1989)]{danz89}. The different behaviour of these two bands may indicate that they arise from different zones in the ejecta. 
\begin{table}
  \begin{center}
  \caption{A summary of current CO and SiO observations in Type II supernovae. The {\bf --} sign refers to a non-detection of the molecular species.}
  \label{tab1}
 {\scriptsize
  \begin{tabular}{lcccl}\hline 
{\bf Name} & {\bf CO 1st overtone } & {\bf CO fundamental } & {\bf SiO fundamental } & {\bf Reference} \\ 
 & [2.3 $\mu$m] & [4.65 $\mu$m] & [8.1 $\mu$m] & \\
\hline
SN1987A& {\bf X} &  & & \cite{spy88} \\
 & &{\bf X} & &\cite{cat88} \\ 
 & & &{\bf X}& \cite{ait88} \\
 &{\bf X}&{\bf X} & & \cite{danz89} \\
& & &{\bf X} &   \cite{roche91} \\
\hline
SN1995ad &{\bf X} & & &\cite{spy96}\\
\hline
SN1998S & {\bf X} & & &\cite{ger00} \\
&{\bf X}& & &  \cite{fas01} \\
\hline
SN1998dl& {\bf X} & & &\cite{spy01} \\
\hline
SN1999em & {\bf X} & & &\cite{spy01} \\
\hline
SN2002hh & {\bf X} & & &\cite{poz06} \\
\hline
SN2004fj& &{\bf X}& &\cite{kot05} \\
& &{\bf X}&{\bf --}& \cite{sza11} \\
\hline
SN2005af & &{\bf X}&{\bf X}&\cite{kot06} \\
\hline
SN2004et &  &{\bf X} &{\bf X} &\cite{kot09} \\
\hline

  \end{tabular}
  }
 \end{center}
 \end{table}

Silicon monoxide, SiO, has been detected in the mid-IR in three SNe through its fundamental band around 8.1 $\mu$m. Its first detection in SN1987A showed that emission was apparent from day 160 to day 520 \cite[(Aitken et al. 1988]{ait88}, \cite[Roche et al. 1991)]{roche91}. No emission was recorded at later times and this decline coincided with the onset of dust formation at $\sim$ 530 days \cite[(Lucy et al. 1989)]{lucy89}. The depletion of SiO into dust was therefore proposed as a scenario for SiO emission decline. A similar behaviour was observed in SN2004et by \cite{kot09}. 

Other molecular species have been proposed to explain unidentified bands in the IR spectrum of SN1987A. For example, the broad emission feature at 3.88 $\mu$m appearing at late time was proposed to come from carbon monosulphide, CS, by \cite{meik93}. These assignments were mainly made on wavelength coincidence and confirmation by chemical models was difficult (e.g., \cite{liu98} for CS, \cite{gear99} for CO$^+$). As to the unidentified emission bands in the near-IR at 3.4 $\mu$m and 3.53 $\mu$m, \cite{mil92} ascribed the emission to \hhh, while \cite{yan98} modelled the chemistry of  a cool molecular clump where \hhh~could form in large enough amount to satisfactorily reproduce the emission band intensity. However, their model assumed that the composition of the molecular clump comprised both hydrogen, \hhh, CO and OH. As mentioned in Section 1, the scarcity of molecules observed in SN ejecta entails the unlikely presence of microscopic mixed hydrogen with atomic carbon and oxygen in the ejecta, thereby invalidating \hhh~as a potential carrier for these bands. Therefore, the only two molecular species with unambiguous identification in SN ejecta remain CO and SiO, and their detection status is summarised in Table~\ref{tab1}.

Chemical models have been attempted to explain the formation of CO and SiO in the ejecta of  SN1987A. \cite{pet89} and \cite{lepp90} proposed the first chemical models of the O- and C-rich zones including ion-molecule reactions, radiative association processes, and reactions with the fast Compton electrons created from the collisional degrading of \gam. They found that helium could not be microscopically mixed in the different layers because the reactive ion He$^+$ destroyed any molecular species. The prevalent formation processes for molecules were radiative association reactions whilst destruction was provided by collisions with Compton electrons. \cite{liu92} and \cite{gear99} modelled the emission bands of CO while \cite[Liu \& Dalgarno (1994, 1996)]{liu94} modelled the emission bands of SiO. All models assumed similar chemical processes as \cite{lepp90} to derive molecular abundances and masses. Finally, \cite{liu95} explored the impact of molecular cooling in the ejecta as CO and SiO are strong coolants through their ro-vibrational lines. Their study of the O-rich core showed that the kinetic temperature could  indeed drop from 5000 K to 1500 K when CO cooling was introduced in their model. 

Central to these chemical models was the assumption that steady-state was quickly established for the chemistry at early times. However, this assumption was later disproved by \cite{cher08} and \cite{cher09} in their study of the chemistry of primitive SNe in the early universe. Based on a more extended chemical network (see Section 3), they showed that the chemical times of key formation and destruction processes were often shorter than the dynamical time, implying that an active chemistry could take place at time greater than 100 days post-outburst.  
For SN ejecta other than SN1987A, no chemical models have been attempted so far. We therefore present in the next section preliminary results for the non-equilibrium chemical kinetic model of a SN ejecta with a 15 \Ms~progenitor mass of solar metallicity. 

\section{New model for stratified, homogeneous ejecta}

\begin{figure}[b]
\begin{center}
 \includegraphics[width=5.1in]{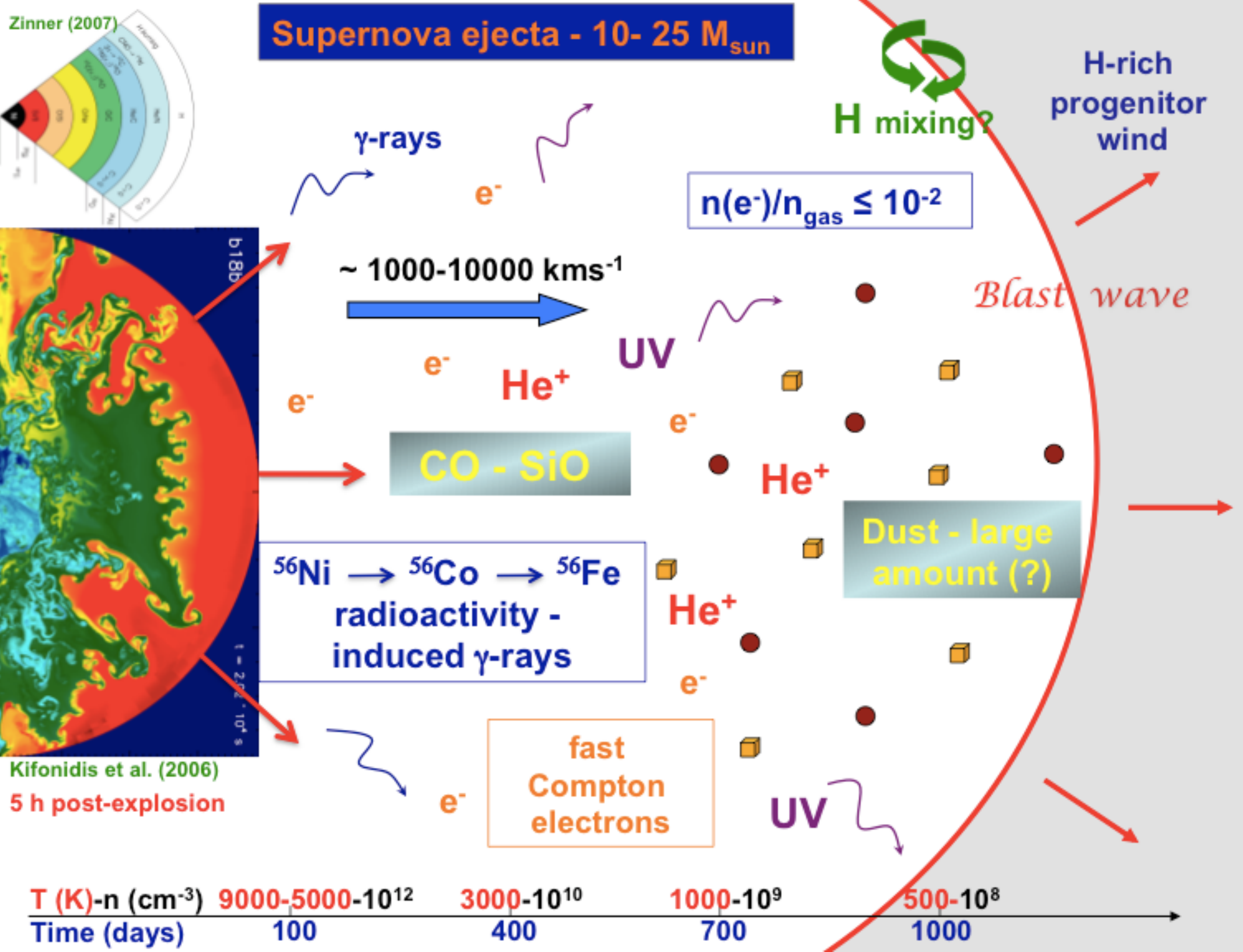} 
 \caption{A schematic view of the ejecta of a Type II supernova. On the upper left-hand corner, the pre-supernova core reflects the various stages of nuclear burning and the stratified lay-out of the heavy-elements in shells (\cite[Zinner 2007]{zin07}). A few hours after the explosion, the helium core experiences Rayleigh-Taylor instabilities, mixing and fragmentation (\cite[Kifonidis et al. 2006]{kif06}). Other quantities are explained in the text.}
\label{fig1}
\end{center}
\end{figure}
\begin{table}
  \begin{center}
  \caption{Mass zones of the helium core for the 15 \Ms~progenitor of solar metallicity. The mass cut is 1.78 \Ms~and the ejected mass is $\sim $ 3.79 \Ms~for a total He-core mass of 5.57 \Ms.  }
  \label{tab2}
 {\scriptsize
  \begin{tabular}{lccc}\hline 
{\bf Zone} & {\bf Position (\Ms)} & {\bf Mass (\Ms)} &{\bf Major constituants} \\ 
\hline
1A & 1.79 - 1.89 & 0.1 &Si-S-Fe  \\
1B & 1.89 - 1.98 & 0.09 &Si-O  \\
2 & 1.98 - 2.27 & 0.29 & O-Si-Mg  \\
3 & 2.27 - 2.62 & 0.35 & O-Mg-Si  \\
4A & 2.62 - 2.81 & 0.19 & O-C  \\
4B & 2.81 - 3.04 & 0.23 &O-C-He  \\
5 & 3.04 - 3.79 & 0.75& He-C-N  \\
\hline

  \end{tabular}
  }
 \end{center}
 \end{table}

We study the chemistry of the ejecta associated with a 15 \Ms~progenitor of solar metallicity. A schematic view of conditions in the ejecta is given in Fig.\,\ref{fig1}. A pre-supernova core explodes and experiences the passage of a blast wave which triggers a reverse shock at the base of the hydrogen envelope. This shock propagates inward and leaves a reversed pressure-gradient in its wake that produces Rayleigh-Taylor instabilities and macroscopic mixing in the helium core. The mixing has ceased after a few days (\cite{jog10}) and the partial fragmentation of the helium core proceeds with time. Radioactive \Ni~decays into \Co~on short time scales, and the later decay of \Co~in~\Fe~creates \gam~that propagate in the ejecta. These \gam~degrade to X-rays and ultraviolet (UV) photons by Compton scattering and create a population of fast Compton electrons in the ejecta. These Compton electrons ionise the  elements, including helium. Typical temperatures and densities are also indicated as a function of time in the ejecta. 

Following \cite{cher09}, the ejecta is assumed homogeneous and consists of several mass zones characterised by an initial chemical composition taken from \cite{rau02} for their study of a 15 \Ms~progenitor of solar metallicity. The elements in each zone are assumed to be microscopically mixed but no elemental mixing between zones is considered. We assume that the helium core is hydrogen-free as no microscopic mixing of hydrogen takes place inward in the core. Hydrogen thus stays confined to the most external layer that corresponds to the progenitor envelope. The zoning considered in this study is briefly summarised in Table \ref{tab2}. We model the formation of several key molecules, including CO, SiO, O$_2$, SO and SiS, as well as small molecular clusters of various solids (e.g., SiO$_2$, FeS, pure metals, carbon chains and rings) as in \cite {cher10}. We study the chemical processes active after outburst from 100 days to 1000 days, a time span relevant for the study of molecule and dust synthesis in the ejecta. They include the formation of molecules by termolcular reactions and by bimolecular processes like neutral-neutral reactions, radiative association reactions and ion-molecule reactions. The destruction processes include ionisation and dissociation processes by Compton electrons and the UV radiation field, ion-molecule reactions, neutral-neutral processes and thermal fragmentation. 

\section{Results on molecules}

\begin{figure}[b]
\begin{center}
 \includegraphics[width=5.1in]{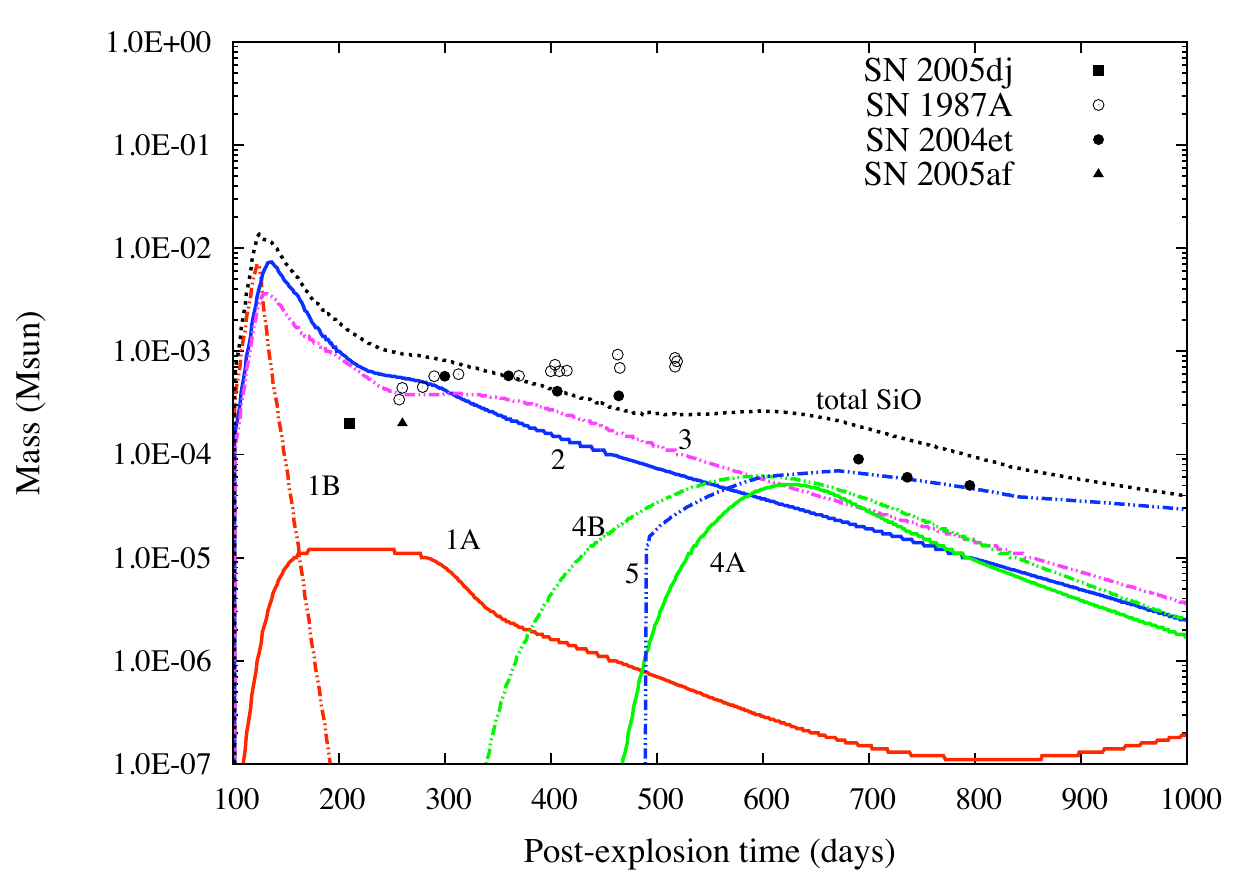} 
 \caption{The mass of SiO as a function of post-explosion time for the various helium core mass zones (see text). SiO masses derived from observational data for several SNe are also shown. }
\label{fig2}
\end{center}
\end{figure}

\begin{figure}[b]
\begin{center}
 \includegraphics[width=5.1in]{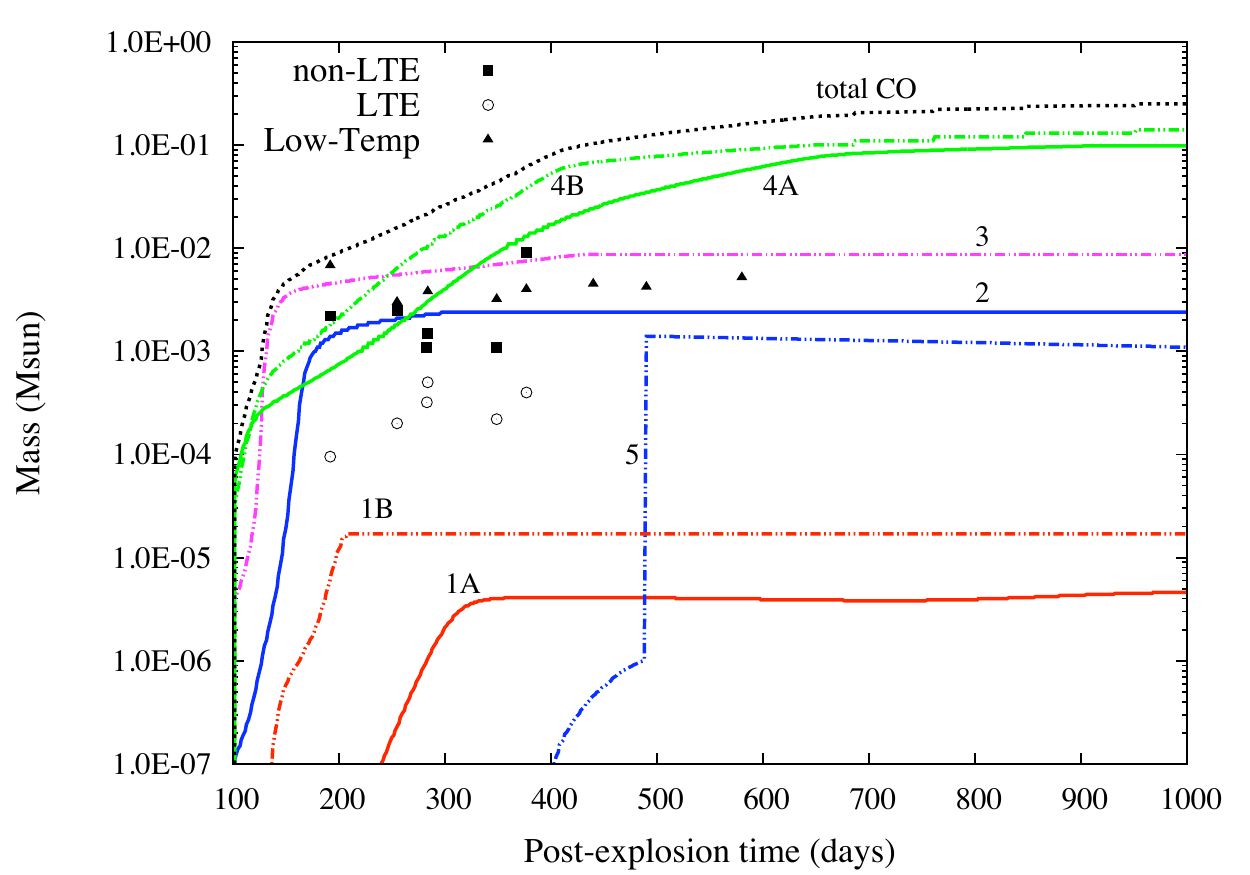} 
 \caption{The mass of CO as a function of post-explosion time for the various helium core mass zones (see text). CO masses derived by \cite{liu92} and \cite{liu95} from observational data of SN1987A are shown.}
\label{fig3}
\end{center}
\end{figure}


The mass of SiO formed in the ejecta as a function of time is illustrated in Figure \ref{fig2}. For comparison, we have included the SiO masses derived from observations of various Type II SNe. Zones 1B, 2 and 3 are the prevalent providers of SiO through formation processes including neutral-neutral and radiative association reactions. The SiO mass variation with time agrees well with the decline of the observed SiO mass, in particular that of SN2004et (Kotak et al. 2009). However, this agreement implies that the effective and steady conversion of gas-phase SiO into small molecular clusters of SiO$_2$ and Si proceeds at early time ($\sim$ 150 days post-outburst) for the three zones where SiO efficiently forms. In this regard, the SiO molecule is a direct tracer of the formation of silicate dust in SN ejecta. Observations of SiO at several post-explosion times thus probe the effectiveness of dust synthesis in these environments. 

Results for CO as a function of mass zones and time are shown in Figure \ref{fig3}. CO masses derived for SN1987A from models by \cite{liu92} and \cite{liu95} are also plotted for comparison. However, a 15 \Ms~progenitor model may not be appropriate for SN1987A, for which a more massive progenitor has been proposed (\cite{woo88}). The metallicity of SN1987A progenitor is also sub-solar and similar to that of the LMC. The trends of the CO masses derived for SN1987A, i.e., an effective formation of CO in large amounts at very early times, are reproduced by our results although we produce more CO in the 15 \Ms~ejecta. CO is prevalently formed in the two outer zones 4A and 4B where oxygen is still overabundant with respect to carbon and where helium is absent (zone 4A) or exists in small amounts (zone 4B). From our results, there exits no direct link between the presence of CO in the ejecta and carbon dust. The latter is traced by the formation of carbon chains and rings in our model, and forms in the outer, He/C-rich zone 5, when the \hei~abundance has diminished, either due to its effective recombination with thermal electrons or to non-effective ionisation of helium by Compton electrons. Small amounts of CO are formed in zone 5 but most of the carbon is either in atomic form or in carbon dust precursors. Therefore, CO formation is not linked to the condensation of carbon solids and CO is not a tracer of carbon dust in SN ejecta. CO is just the prevalent molecule that forms in zones rich in oxygen and carbon, and poor in helium, whatever the initial C/O ratio of the zones.  

Other molecular species are produced in large amounts in the ejecta and trace the specific chemical composition of their formation zones. In the inner most mass zone 1A, silicon sulphide, SiS, forms efficiently from radiative association and neutral-neutral reactions, pumping about 34 \% of the atomic Si present in the zone. The SiS mass at day 1000 is $ \sim 3 \times 10^{-2}$ \Ms. Because zones 2 and 3 are very rich in free atomic oxygen, the formation of dioxygen, O$_2$ proceeds as early as 300 days. In zone 1B and 2, sulphur is still abundant and its reaction with O$_2$ induces the formation of sulphur monoxide, SO, as soon as O$_2$ forms. The final masses of O$_2$ and SO at 1000 days post-explosion are $\sim 3 \times 10^{-1}$ \Ms~and $ \sim 1.5 \times 10^{-2}$ \Ms, respectively.  

\section{Discussion}

The ejecta of SNe form two types of molecular species, those directly involved in the formation processes of dust (e.g., SiO) and those only depleting gas-phase elements (e.g., CO, O$_2$).  Members of the first category are rapidly depleted in the synthesis of dust  and are thus direct tracers of the dust formation processes. Therefore, dedicated observational surveys of these specific species at various times after outburst would be extremely useful to directly assess the efficiency with which the SN ejecta produces dust. Molecules of the second kind are of great interest as well. For the present 15 \Ms~progenitor model with solar metallicity, they represent $\sim$ 17\% of the ejected mass at day 1000 and include CO, O$_2$, SiS and SO. Somewhat larger mass fractions of ejected molecules were derived for zero-metallicity SNe in the early universe (Cherchneff \& Dwek 2009). These large molecular fractions impact the gas in two major ways. Firstly, the depletion by molecules of elements entering the processes of dust synthesis takes place and needs to be considered when assessing the amount of dust formed in SNe. This is well exemplified by the formation of gas-phase S$_2$ and SiS in zone 1A, which depletes atomic sulphur, thus impacting on the formation of FeS clusters. Secondly, molecules like CO are strong coolants through their ro-vibrational transitions and their presence will affect the thermal history of the environment where they reside. The cooling provided by CO molecules in O/C-rich clumps applied to the ejecta of SN1987A was addressed by \cite{liu95}. They considered clumps with equal masses of atomic O and C and where 1\% of the mass was in CO.  The cooling provided by CO could decrease the gas kinetic temperature down by a factor of three compared to CO-free gas where cooling was essentially provided by the 6300 {\AA} and 6363 {\AA} forbidden lines of [OI]. In the present model, the CO mass corresponds to $\sim$ 2\% of the O-rich core mass (i.e., zones 2 and 3) at day 400, where the mass fraction reaches $\sim 24$\% of the CO formation zone (i.e., zones 4A and 4B) at a similar time. The cooling provided by CO molecules must thus be strong as derived by \cite{liu95}. Such a cooling will favour more efficient molecule and dust synthesis in the ejecta dense clumps. 

Species of the second category are also crucial to the study of the SN ejecta evolution at very late time, i.e., in young SN remnants like Cas A. Molecules like CO or SiS are expected to survive in the dense, cool clumps where they formed and young remnants that have not yet been shocked by the reverse shock are expected to bear such a molecular signature. These species should thus be detectable at sub-millimetre wavelengths in young SN remnants and would not only provide clues on the local kinetics and the chemistry active in clumps or filaments but also on their chemical synthesis in the prior SN stage.   
\bigskip
\acknowledgement
This research is supported by the Swiss National Science Foundation grant No 20GN21-128950 through the European Science Foundation Collaborative Research Project CoDustMas.

\begin{discussion}

\discuss{Posch}{How efficient do you expect dust formation by supernovae to be in the end, considering dust destruction in supernova remnants?}

\discuss{Cherchneff}{My impression is that they are efficient dust makers in the end. But an answer to your question requires accurate models of dust formation in supernovae and the impact of the reverse shock on dust in the remnant. Those models do not yet exist. Existing studies consider the classical nucleation theory to model dust formation which is not appropriate for the case of a supernova ejecta. The fragmentation and clumpiness of the medium must also be taken into account. In the remnant, the impact of the reverse shock on ejecta dust has also been studied, but assuming a homogenous ejecta. Destruction efficiencies vary between 70\% and 99\% of the initial dust mass depending on the pre-shock gas density. So a very quick assessment of the dust mass surviving in the remnant would be between 0.001 \Ms~and 0.03~\Ms~assuming an average dust mass of 0.1~\Ms~synthesised by a 20~\Ms~progenitor. These values are comparable to the amount of dust produced by a carbon star over its lifetime. Furthermore, recent observations of Cas A with Spitzer and Herschel have detected warm and cool dust in the remnant. And supernova dust is extracted from meteorites. So somehow, supernovae form dust and some of it survives in the remnant and continues its journey to the interstellar medium.  }

\end{discussion}


\begin{thebibliography}{}

\bibitem[Aitken et al. (1988)]{ait88}
{Aitken, D.K.., Smith, C.H., James, S.D. et al.} 1988,
\textit{M.N.R.A.S.}, 235, 19

\bibitem[Catchpole et al. (1988)]{cat88}
{Catchpole, R.M., Whitelock, P.A., Feast M.W. et al.} 1988,
\textit{M.N.R.A.S.}, 231, 75

\bibitem[Cherchneff (2006)]{cher06}{Cherchneff, I.} 2006, 
 \textit{A\&A} 456, 1001

 \bibitem[Cherchneff \& Lilly (2008)]{cher08}{Cherchneff, I. \& Lilly,, S.} 2008, 
 \textit{ApJ} 683, L123

 \bibitem[Cherchneff \& Dwek (2009)]{cher09}{Cherchneff, I. \& Dwek,, E.} 2009, 
 \textit{ApJ} 703, 642
 
 \bibitem[Cherchneff \& Dwek (2010)]{cher10}{Cherchneff, I. \& Dwek,, E.} 2010, 
 \textit{ApJ} 713, 1

\bibitem[Danziger et al. (1988)]{danz88}
{Danziger, I.J., Bouchet, P. Fosbury, R.A.E. et al.} 1988, in: 
 \textit{Supernova 1987A in the Large Magellanic Cloud} (Cambridge and New York: Cambridge University Press), p.\,37
 
\bibitem[Danziger et al. (1989)]{danz89}
{Danziger, I.J., Lucy, L.B., Bouchet, P. et al.} 1989, in: 
 \textit{Supernova 1987A in the Large Magellanic Cloud} (Cambridge and New York: Cambridge University Press), p.\,37
 
 \bibitem[Fassia et al. (2001)]{fas01}{Fassia, A., Meikle, P., Chugai, N. et al.} 2001, 
 \textit{M.N.R.A.S}, 325, 907
 
 \bibitem[Gearhart et al. (1999)]{gear99} {Gearhart, R.A., Wheeler, J.C. \& Swartz, D.A. }1999, 
  \textit{ApJ}, 510, 944
 
 \bibitem[Gerardy et al. (2000)]{ger00}{Gerardy, C.L., Fesen, R.A., Hoeflich, P. \& Wheeler, J.C.} 2000, 
 \textit{AJ} 119, 2968
 
  \bibitem[Joggerst et al. 2010]{jog10}{Joggerst, C.C., Almgren, A., \& Woosley, S.E.} 2010, 
 \textit{ApJ} 723, 353
 
 \bibitem[Kifonidis et al. (2006)]{kif06}{Kifonidis, K. Plewa, T., Scheck, L. et al.} 2006, 
 \textit{A\&A} 463, 661
 
 \bibitem[Kotak et al. (2005)]{kot05}{Kotak, R., Meikle, P., van Dyck, S. et al. } 2005, 
 \textit{ApJ} 628, L123

 
 \bibitem[Kotak et al. (2006)]{kot06}{Kotak, R., Meikle, P., Pozzo, M. et al. } 2006, 
 \textit{ApJ} 651, L117
 
 \bibitem[Kotak et al. (2009)]{kot09}{Kotak, R., Meikle, P., Farrah, D. et al. } 2009, 
 \textit{ApJ} 704, 306
 
\bibitem[Lepp et al. (1990)]{lepp90} {Lepp, S., Dalgarno, A., \& McCray, R.} 1990, 
\textit{ApJ}, 358, 262

\bibitem[Liu et al. (1992)]{liu92} {liu, W., Dalgarno, A., \& Lepp, S.} 1992, 
\textit{ApJ}, 396, 679

\bibitem[Liu \& Dalgarno (1994)]{liu94} {Liu, W. \& Dalgarno, A. }1994, 
\textit{ApJ}, 438, 789

\bibitem[Liu \& Dalgarno (1995)]{liu95} {Liu, W. \& Dalgarno, A.} 1995, 
\textit{ApJ}, 454, 472

\bibitem[Liu \& Dalgarno (1996)]{liu96} {Liu, W. \& Dalgarno, A. }1996,
\textit{ApJ}, 471, 480

\bibitem[Liu (1998)]{liu98} {Liu, W.}1998,
\textit{ApJ}, 496, 967
 
 \bibitem[Lucy et al. (1989)]{lucy89}
{Lucy, L.B., Danziger, I.J., Gouiffes, C. \& Bouchet, P.} 1989, in: 
 \textit{Supernova 1987A in the Large Magellanic Cloud} (Cambridge and New York: Cambridge University Press), p.\,37
 
\bibitem[Meikle et al. (1993)]{meik93}{Meikle, P., Spyromilio, J., Allen, D.A. et al.} 1993,
 \textit{M.N.R.A.S.} 261, 235
 
 \bibitem[Miller et al. (1992)]{mil92}{Miller, S., Tennyson, J.,  Lepp, S. \& Dalgarno, A.} 1992, 
 \textit{Nature}, 355, 420
 
 \bibitem[Petuchowski et al.  (1989)]{pet89} {Petuchowski, S.J., Dwek, E., Allen, J.E. \& Nuth III, J.A. }1989,
\textit{ApJ}, 342, 406
 
 \bibitem[Pozzo et al. (2006)]{poz06}{Pozzo, M., Meikle, Rayner, T.J. et al.}  2006, 
 \textit{M.N.R.A.S}, 368, 1169
 
 \bibitem[Rauscher et al. (2002)]{rau02}{Rauscher, T., Heger, A., Hoffman, R.D. \& Woosley, S.E.} 2002, 
 \textit{ApJ}, 576, 323

 \bibitem[Roche et al. (1991)]{roche91}{Roche, P.F., Aitken, D.K., \& Smith, C.H.} 1991, 
 \textit{M.N.R.A.S}, 252, 39
 
 \bibitem[Spyromilio et al. (1988)]{spy88}{Spyromilio, J., Meikle, W.P.S.,  Learner, R.C.M. \& Allen, D.A.} 1988, 
 \textit{Nature}, 334, 327
 
 \bibitem[Spyromilio \& Leibundgut (1996)]{spy96}{Spyromilio, J. \& Leibundgut , B.} 1996, 
 \textit{M.N.R.A.S}, 283, L89
 
 \bibitem[Spyromilio et al. (2001)]{spy01}{Spyromilio, J., Leibundgut , B. \& Gilmozzi, R.} 2001, 
 \textit{A\&A}, 376, 188
 
 \bibitem[Szalai et al. (2011)]{sza11}{Szalai. T, Vink{\'o}, Balog, Z. et al.} 2011, 
 \textit{A\&A}, 527, 61
 
 \bibitem[Woosley 1988]{woo88}{Woosley, S.E.} 1988, 
 \textit{ApJ} 218, 253
 
\bibitem[Yan \& Dalgarno (1998)]{yan98} {Yan, Min. \& Dalgarno, A.} 1998, 
\textit{ApJ}, 500, 1049

 \bibitem[Zinner (2007)]{zin07}
{Zinner, E.} 2007, in: K.K. Turekian, H.D. Holland \& A.M. Davis (eds.), 
 \textit{Treatise in Geochemistry 1} (Oxford and San Diego: Elsevier), p.\,1

\end{thebibliography}
\end{document}